\documentclass[12pt]{elsarticle}
\usepackage{natbib,amsmath}

\begin{document}

\def\nhdlls{154}
\def\nsource{403277}  
\def\nspectra{434686}  
\def\npair{179}  
\def\nlee{several thousands}
\def\hub{h_{72}^{-1}}
\def\umfp{{\hub \, \rm Mpc}}
\def\mew{W_\lambda}
\def\ew{$\mew$}
\def\mzq{z_q}
\def\mfmin{f_{\rm min}}
\def\fmin{$\mfmin$}
\def\zabs{$z_{\rm abs}$}
\def\mzabs{z_{\rm abs}}
\def\msna{{\rm S/N}^{\rm A}_{912}}
\def\sna{S/N$^{\rm A}_{912}$}
\def\mnull{\nu_{\rm 912}}
\def\nnull{$\nu_{\rm 912}$}
\def\intl{\int\limits}
\def\nstatqso{193}
\def\maxoff{0.4}
\def\clls{1.9 \pm 0.2}
\def\alls{5.2 \pm 1.5}
\def\blls{-0.9^{+0.4}_{-0.05}}
\def\cmma{\;\;\; ,}
\def\perd{\;\;\; .}
\def\ltk{\left [ \,}
\def\ltp{\left ( \,}
\def\ltb{\left \{ \,}
\def\rtk{\, \right  ] }
\def\rtp{\, \right  ) }
\def\rtb{\, \right \} }
\def\sci#1{{\; \times \; 10^{#1}}}
\def \rAA {\rm \AA}
\def \mzem {z_{\rm em}}
\def\smm{\sum\limits}
\def \cmm  {cm$^{-2}$}
\def \cmmm {cm$^{-3}$}
\def \kms  {km~s$^{-1}$}
\def \mkms  {{\rm km~s^{-1}}}
\def \lyaf {Ly$\alpha$ forest}
\def \Lya  {Ly$\alpha$}
\def \lya  {Ly$\alpha$}
\def \mlya  {Ly\alpha}
\def \Lyb  {Ly$\beta$}
\def \nhi  {$N_{\rm HI}$}
\def \mnhi  {N_{\rm HI}}
\def \lnhi {$\log N_{HI}$}
\def \mlnhi {\log N_{HI}}
\def \etal {\textit{et al.}}
\def \ob {$\Omega_b$}
\def \obh {$\Omega_bh^{-2}$}
\def \om {$\Omega_m$}
\def \ol {$\Omega_{\Lambda}$}
\def \gz {$g(z)$}
\def \mgz {g(z)}
\def \lyaf {Lyman--$\alpha$ forest}
\def \fnhi {$f(\mnhi,X)$}
\def \mfnhi {f(\mnhi,X)}
\def \mnmin {\mnhi^{\rm min}}
\def \nmin {$\mnhi^{\rm min}$}
\newcommand{\cm}[1]{\, {\rm cm^{#1}}}
\def\N#1{{N({\rm #1})}}
\def\psol#1#2#3#4{$\{ {\rm #1}^{#2}/{\rm #3}^{#4}\}$}
\def\pxh{$\{ {\rm X/H} \}$}
\def \snrlim {SNR$_{lim}$}
\def\mglls {\gamma_{\rm LLS}}
\def\mavgt {<\mtll>}

\def\aap{A \& A}
\def\aj{AJ}
\def\apj{ApJ}
\def\apss{Ap\&SS}
\def\apjl{ApJL}
\def\apjs{ApJS}
\def\apjsupp{ApJS}
\def\araa{ARAA}
\def\mnras{MNRAS}
\def\nat{Nature}
\def\pasp{PASP}
\def\prd{PhRvD}
\def\aapr{A\&A Rev.}
\def\physrep{Physics Reports}
\def\nar{New A Rev.}
\def\aaps{A\&AS}

\begin{frontmatter}

\title{The igmspec Database of Public Spectra Probing
the Intergalactic Medium}


\author{
J. Xavier Prochaska}
\address{
Department of Astronomy and Astrophysics, UCO/Lick Observatory, University of California, 1156 High Street, Santa Cruz, CA 95064}

\begin{abstract}
We describe v02 of {\it igmspec}, a database of publically
available ultraviolet, optical, and near-infrared spectra 
that probe the intergalactic medium (IGM).  This database, a child
of the {\it specdb} repository in the {\it specdb} github organization, 
comprises \nsource~unique
sources and \nspectra~spectra obtained with the world's greatest
observatories.  All of these data are distributed in a single
$\approx 25$\,GB HDF5 file maintained at the University of
California Observatories and the University of California,
Santa Cruz.  The {\it specdb} software package includes
Python scripts and modules for searching the source catalog
and spectral datasets, and software links to the {\it linetools}
package for spectral analysis.
The repository also includes software to generate 
private spectral datasets that are compliant 
with International Virtual Observatory Alliance (IVOA) protocols
and a Python-based interface for IVOA Simple Spectral Access
queries.
Future versions of {\it igmspec} will ingest other sources
(e.g.\ gamma-ray burst afterglows) and other surveys as they become
publicly available.  
The overall goal is to include every 
spectrum that effectively probes the IGM.  
Future databases of {\it specdb}
may include 
publicly available galaxy spectra 
({\it exgalspec}) and 
published supernovae spectra ({\it snspec}). 
The community is encouraged to join the effort on github:
https://github.com/specdb
\end{abstract}

\end{frontmatter}


\section{Introduction}
\label{sec:intro}

Shortly after the discovery of quasars \cite{schmidt63},
astronomers identified absorption-line features in their 
spectra indicating the presence of intergalactic gas
\citep{bs65,blb66}.  
The community quickly recognized both the value of these
data to cosmology and galaxy formation \citep{gp65,bs69}.
In the decades that followed, observationalists gathered
spectra of increasingly higher quality on several tens of
high-$z$ sources to analyze the IGM 
\citep{sargent80,tytler82,wolfe86,lzt91}.
These were obtained primarily with private observatories and
the spectra were rarely made available
to the community in science-ready form.
An obvious exception was the data archives of the
International Ultraviolet Explorer
and the Hubble Space Telescope 
\citep[{\it HST};][]{hstkeyproj_1,bechtold02},
but even these have been difficult to  collate and combine.
One also recognizes the policy of European Southern
Observatory to archive and
make public their Very Large Telescope (VLT) datasets.

The past $\approx 10$\,years has witnessed the
rise of large, public spectral datasets, especially
the 2dF Survey and Sloan Digital Sky Survey \citep[SDSS;][]{yaa+00,croom01}.
These include the spectra of several hundred thousand quasars
that probe the IGM \citep[e.g.][]{sdss_qso_dr7}.
Their surveys have further stimulated the public release
of smaller spectroscopic surveys with higher quality
data (S/N, resolution) on complimentary sources
\citep[e.g.][]{pwh+07,prochaska+15}.
Accessing both the large survey datasets 
and these modest high-quality datasets of IGM spectroscopy
has remained a challenge, however. 
This is primarily due to the diverse range of 
data formats adopted within 
the astronomical community.
For example, science-grade spectra are often processed and combined
with custom software that may not preserve or record salient meta 
data.  While the International Virtual Observatory Alliance (IVOA)
has taken significant effort to establish a
spectral data model\footnote{http://www.ivoa.net/documents/SpectralDM/}
and  standards for Simple Spectral Access\footnote{http://www.ivoa.net/documents/SSA/} (
SSA) protocols, these have seen limited usage.
Despite the fact that the entire set of
reduced and calibrated spectra for the IGM
comprises less than a few tens GB of disk space,
there has been no method established for
wide-spread distribution.

Therefore, as a service to IGM researchers, ourselves, and the
broader community, we have initiated an effort to collate 
all of the published surveys of IGM spectroscopy. 
These are packaged in a single HDF5 file referred to as
the {\it igmspec} database, staged for direct download
at the University of California, 
Santa Cruz\footnote{http://specdb.ucsc.edu}.
We also have developed Python software 
for querying the source catalog and meta data
and for accessing the spectra. 
This includes an SSA compliant interface for data queries,
although we do not yet include a web portal for SSA service.

The uber-project of {\it igmspec}
is {\it specdb}\footnote{http://specdb.ucolick.org}, 
which includes a suite of software for the generation 
and manipulation  of spectral databases.
With this publication we provide the first database,
{\it igmspec}, which focuses on spectra that probe the IGM. 
In v02\footnote{v01 of {\it igmspec} was a prototype and was
accessed by only a few guinea pigs.}, 
we consider only quasars but 
this database will grow with future surveys 
and we will be expanded to include other sources
\citep[e.g. gamma-ray burst afterglow spectra, star-forming
galaxies, supernovae;][]{fjp+09,rpk+10,cooke+12}
We will also ingest other historical 
datasets that come to light in
the public domain.  
In {\it specdb}, we also provide 
software to generate private spectral databases; 
this may be of interest for surveys 
under construction, i.e.\ prior to publication.

This paper describes version v02 of the {\it igmspec}
database and an overview of the {\it specdb} software.  
Additional documentation is available\footnote{http://specdb.readthedocs.io}
on {\it Read the Docs}.
The manuscript is organized as follows:
Section~\ref{sec:catalog} describes the catalogs
included in {\it igmspec}.
Section~\ref{sec:datasets} details the spectral datasets
in v02 of {\it igmspec}.
Sections~\ref{sec:arch} and \ref{sec:software}
briefly describe the database architecture and
related software.

\section{The {\it igmspec} Source Catalog}
\label{sec:catalog}

At the heart of {\it igmspec}, and any other database  
of {\it specdb}, is a catalog of unique sources.
Each source is assigned an identifier or IDKEY
(e.g.\ IGM\_ID for {\it igmspec})
to be preserved in all future versions of the database.
If a source is discovered to be erroneous, 
we will remove it without modifying
any other IGM\_ID values.

To construct the source catalog of {\it igmspec}, 
we followed these steps, in order:

\begin{enumerate}
\item Ingest all quasars from the BOSS DR12 survey.\footnote{See
http://www.sdss.org/dr12/algorithms/boss-dr12-quasar-catalog/}
\item Add all quasars from the SDSS DR7 survey not observed by BOSS.
These were defined as any sources with $\theta > 2''$ angular separation
from sources in the BOSS dataset.
\item Add any additional, unique sources ($\theta > 2''$)
from the {\it igmspec} datasets.
\item Examine each pair of sources with $\theta \le 10''$ 
(\npair\ total) to establish whether they are truly unique.
\end{enumerate}
The {\it igmspec} source catalog includes a minimum of meta data
to limit its size and maximize search speed.
The columns are described in Table~\ref{tab:cat_keys}.
We caution that other than the BOSS and SDSS surveys 
the adopted astrometry may not have sub-arcsecond accuracy.

\begin{table}
\caption{CATALOG META DATA\label{tab:cat_keys}}
\footnotesize
\begin{tabular}{lcl}
Column & Type  & Description \\
\hline
RA           & float64 & Right Ascension in J2000 (degrees) \\
DEC          & float64 & Declination in J2000 (degrees) \\
IGM\_ID      & int     & Unique {\it igmspec} identifier \\
flavor       & str     & Type of source (quasar, GRB, galaxy) \\
zem          & float64 & Redshift of the source \\
sig\_zem     & float64 & Uncertainty in the source redshift \\
flag\_zem    & str     & String describing the redshift measurement \\
flag\_group  & int     & Bitwise flag indicating the data groups covering the source \\
\hline
\end{tabular}
\end{table}


\section{{\it igmspec} Data Groups}
\label{sec:datasets}

This manuscript describes v02 of the {\it igmspec}
database and represents our first comprehensive effort to collate
spectra probing the IGM at UV and optical wavebands at all
redshifts.  
The first version was a prototype and is not being
distributed.
Future efforts will extend to other wavebands and/or
additional types of sources 
\citep[e.g. star-forming galaxies;][]{rpk+10}.

Table~\ref{tab:datasets} lists the surveys included in
v02 of {\it igmspec} and summarizes properties
of each survey.  
Within our database terminology, each of these
is referred to as a data group.
This table lists the number of unique sources and spectra
ingested.  We have not explicitly culled any data
from the survey, e.g. by S/N or any other attribute.
There were occasional spectra in the surveys
that identified as junk and ignored (see details in the
sub-sections). 

Table~\ref{tab:meta_spec} lists the meta data included
for every spectrum ingested into {\it igmspec}.
The RA\_GROUP, DEC\_GROUP, and zem\_GROUP values are the
coordinates and redshifts for the source as reported by
the survey.  As described in Section~\ref{sec:catalog},
these have been collated across all data groups for the
database to generate a catalog of unique sources.
Therefore, the values in the meta tables need not 
identically match those in the source catalog.
In addition, each meta table includes a title, reference,
and additional data descriptions (in HDF5 attributes)
to be compliant with the IVOA spectral data model
and to enable compliant SSA queries.
Each data group also typically includes additional meta data
specific to it (e.g.\ photometry for SDSS sources).
The {\it igmspec} documentation and 
the original references provide greater detail on the meta data
and spectra included within each survey, but the following
sub-sections also provide brief descriptions. 

The majority of IGM research is performed on continuum-normalized
spectra to assess the opacity of intervening gas.
Table~\ref{tab:datasets} indicates whether the spectral flux
is normalized or, for unnormalized data, whether the flux calibration
was relative or absolute.  For many of the unnormalized
datasets, we provide a separate estimate of the 
quasar continuum in the database that can be applied in
software for normalization.
These continua were primarily generated from low-order
polynomial fits to the spectrum or `by-hand' with spline models
of unabsorbed regions but
see the following sub-sections for further details.
We caution that is limited overlap between
datasets, i.e.\ nearly the same spectrum has been
ingested twice in multiple data groups.  Such
duplications can be identified by
referring to the observation date (DATE-OBS). 

\clearpage
\begin{table}[ht]
\caption{{\it igmspec} DATA GROUPS \label{tab:datasets}}
\begin{tabular}{lcccccc}
Group & $N_{\rm source}^a$ 
& $N_{\rm spec}^b$ & $\lambda_{\rm min}$ (\AA) 
& $\lambda_{\rm max}$ (\AA) & $R^c$  & Flux$^d$\\ 
\hline 
2QZ& 23539& 23539& 3554& 8076& 580& RELATIVE\\ 
BOSS\_DR12& 302257& 302323& 3545& 10414& 2100& ABSOLUTE\\ 
COS-Dwarfs& 43& 43& 1135& 1796& 20000& ABSOLUTE\\ 
COS-Halos& 38& 70& 1135& 5896& 20000& MIXED\\ 
ESI\_DLA& 87& 87& 3993& 10136& 6060& RELATIVE\\ 
GGG& 163& 326& 4317& 10299& 886& RELATIVE\\ 
HD-LLS\_DR1& 127& 145& 3027& 11715& 25000& NORMALIZED\\ 
HDLA100& 86& 86& 3055& 10029& 48000& NORMALIZED\\ 
HSTQSO& 762& 904& 1126& 3302& 14000& ABSOLUTE\\ 
HST\_z2& 69& 69& 1648& 9867& 70& ABSOLUTE\\ 
KODIAQ\_DR1& 170& 235& 2995& 9725& 48000& NORMALIZED\\ 
MUSoDLA& 88& 94& 2989& 10492& 4225& NORMALIZED\\ 
SDSS\_DR7& 105783& 105783& 3782& 9266& 2000& ABSOLUTE\\ 
UVES\_Dall& 40& 40& 3042& 10091& 45000& RELATIVE\\ 
UVpSM4& 69& 642& 0& 10259& 20000& RELATIVE\\ 
XQ-100& 100& 300& 3100& 24803& 5300& RELATIVE\\ 
\hline 
\multicolumn{6}{l}{{$^a$}{Number of unique sources in the dataset. }} \\ 
\multicolumn{6}{l}{{$^b$}{Number of unique spectra in the dataset. }} \\ 
\multicolumn{6}{l}{{$^c$}{Characteristic FWHM resolution of the spectra. }} \\ 
\multicolumn{6}{l}{{$^d$}{Indicates whether the data are fluxed (absolute or relative) or normalized. The COS-Halos spectra include both fluxed (COS) and normalized (HIRES) spectra.}} \\ 
\end{tabular} 
\end{table}

\begin{table}[ht]
\caption{DATASET META DATA\label{tab:meta_spec}}
\footnotesize
\begin{tabular}{lcl}
Column & Type  & Description \\
\hline
RA\_GROUP    & float64 & Right Ascension in J2000 (degrees) \\
DEC\_GROUP   & float64 & Declination in J2000 (degrees) \\
EPOCH        & float64 & Epoch \\
zem\_GROUP   & float64 & Redshift of the source \\
flag\_zem    & str     & Description of the redshift source \\
IGM\_ID      & int     & Unique {\it igmspec} identifier \\
GROUP\_ID    & int     & Unique group identifier \\
DISPERSER    & str     & Name of the dispersing element used \\
INSTR        & str     & Name of the instrument used \\
TELESCOPE    & str     & Name of the telescope used \\
DATE-OBS     & str     & Date of observation (YYYY-MM-DD) \\
SPEC\_FILE   & str     & Name of individual file containing the spectrum \\
R            & float64 & Spectral resolution (FWHM) \\
WVMIN        & float64 & Minimum wavelength of the spectrum (\AA) \\
WVMAX        & float64 & Maximum wavelength of the spectrum (\AA) \\
NPIX         & int     & Number of pixels in the spectrum$^a$ \\
\hline
\multicolumn{3}{l}{
{$^a$}{This does not include any pixels that `pad' the
spectrum at the highest and 
}} \\
\multicolumn{3}{l}{lowest wavelengths,  i.e.\ that have sig~$\le 0$.}
\end{tabular}
\end{table}

\subsection{BOSS DR12}

The Baryonic Oscillations Spectroscopic Survey (BOSS)
observed several hundred thousand quasars as part of its
primary survey.  With its final, complete data release
(DR12), the BOSS team provided several catalogs of quasars
observed by the main survey.  We have drawn all sources
from the three catalogs at their main website.\footnote{http://www.sdss.org/dr12/algorithms/boss-dr12-quasar-catalog/}

The BOSS spectra bundled in v02 of {\it igmspec} were pulled
from the main data server and correspond to versions 
v5\_7\_0 or v5\_7\_2 of the data reduction pipeline.
In addition to the calibrated spectra, we include a
continuum estimate for the majority of quasars.  
For wavelengths long-ward of the quasar's \lya\ emission we 
have ingested
the continuum models generated by G. Zhu 
\citep[see][for details on the algorithm]{zhu+14}.
For the \nlee~quasars analyzed to assess
the flux probability distribution function of the 
\lya\ forest \cite{lee+13}, we include their 
mean-flux-regulated continua \citep{lee+12}.

\subsection{SDSS DR7}
\label{sec:dr7}

The Sloan Digital Sky Survey observed over 100,000 quasars
as part of the SDSS-I survey.  These were primarily targeted
based on their optical photometry \citep[e.g.][]{richards09}.  
Upon completion of their final data release (DR7),
the team provided a catalog of quasars \citep{sdss_qso_dr7}.
This forms the basis of the dataset in v02 of {\it igmspec}.
We have also ingested estimates of the quasar continua, as
published in \cite{zhu+14}.  Their analysis focused on a 
separate list of quasars from SDSS DR7 which overlaps the 
\cite{sdss_qso_dr7} catalog, but not completely. 
In a future version, we will provide continua for the majority
of these spectra and also additional quasars discovered in 
SDSS-I but not part of the catalog \cite{sdss_qso_dr7}.



\subsection{2QZ}
\label{sec:2qz}

The 2QZ survey is a catalog of quasars discovered in the 
course of the 2dF redshift survey on the 3.9\,m
Anglo-Australian Telescope \citep{croom01}.  The
majority of spectra are available online\footnote{http://www.2dfquasar.org/Spec\_Cat/2qzsearch2.html}.
We retrieved all available sources in their catalog 
with $z_{\rm em}> 0.05$ and discarded several hundred with
null values in their entire error arrays.

\subsection{KODIAQ\_DR1}
\label{sec:kodiaq}

The Keck Observatory Database of Ionized Absorption toward 
Quasars (KODIAQ) survey is a data release of normalized
quasar spectra obtained with the HIRES spectrometer 
\citep{vogt94} on the Keck~I telescope. 
The first Data Release (DR1) became available in 2015 
\cite{kodiaq_dr1}. 
We have ingested the complete DR1 dataset.

\subsection{HD-LLS DR1}
\label{sec:hdlls}

The high dispersion Lyman Limit System (HD-LLS) sample is a set of 
normalized echelle and echellette spectra obtained with spectrographs
at the Keck and Magellan observatories \cite{prochaska+15}.  
These were primarily acquired
to perform an analysis of $z \sim 3$ LLS. 
The quasars are a heterogeneous set of sources that
are useful  (i.e.\ bright) for such analysis. 

\subsection{GGG}
\label{sec:ggg}

The Giant Gemini GMOS (GGG) survey is a spectroscopic survey of 
$z>4.4$ quasars drawn from the Sloan Digital Sky Survey and re-observed with the GMOS spectrometer on the Gemini North and South telescopes. 
The data release is described in \cite{worseck+14}.

\subsection{XQ-100}
\label{sec:xq100}

The XQ-100 survey is the result of a Large VLT program
titled "Quasars and their absorption lines: 
a legacy survey of the high-redshift universe with VLT/XSHOOTER" 
as described in \cite{xq100}.
The survey comprises XSHOOTER spectra of 100 quasars 
at $z>3.5$ and is the only dataset of v02 in {\it igmspec}
with near-IR coverage.
Note that the coordinates provided in their archival products
are erroneous by up to several arcseconds, 
but we have cross-matched these to the correct sources.
We have also ingested estimate of the continuum for each
source that are provided with the archival spectra.

\subsection{HDLA100}
\label{sec:hdla100}

\cite{marcel13} analyzed a set of 100 representative
damped \lya\ systems
(DLAs) at $z>2$
observed with Keck/HIRES for kinematic and abundance
analyses.  We provide their normalized spectra
\citep[see also][]{pwh+07}.

\subsection{ESIDLA}
\label{sec:esidla}

\cite{rafelski+12,rafelski+14} performed a dedicated survey
with the ESI spectrometer \citep{sbe+02} on the Keck~II telescope
to study $z>4$ DLAs.  This dataset is their
full sample of spectra.

\subsection{MUSoDLA}
\label{sec:musodla}

\cite{jorgenson+13} performed a survey of DLAs drawn from the
SDSS with follow-up spectra obtained primarily with the 
MagE spectrometer at the Magellan Observatory.  The survey was 
supplemented by echelled data taken from the Keck and VLT
archives.  The spectra, as provided in {\it igmspec}, were continuum
normalized by \cite{jorgenson+13}.

\subsection{UVES\_Dall}
\label{sec:uvesdall}

\cite{dww08} processed archival VLT/UVES spectra for 40 quasars 
to study the IGM at $z \sim 2$ with emphasis on the proximity effect.
The quasars were selected to have $z_{\rm em} \approx 2.5$ and the
spectra were required to have high S/N in the \lya\ forest.
We provide the complete dataset of fluxed spectra and also include
an estimate of the quasar continuum.

\subsection{COS-Halos}
\label{sec:cos-halos}

The COS-Halos survey obtained spectra with the
Cosmic Origins Spectrometer 
\citep[COS;][]{cos} on the {\it HST}
to examine the circumgalactic medium (CGM) of luminous,
$z \sim 0.2$ galaxies \citep{tumlinson+13}.
We provide these COS quasar spectra, binned at 3 pixels.
The team also obtained Keck/HIRES spectra for a subset of the
quasars \citep{werk+13}.
All of these data are provided in {\it igmspec}.

\subsection{COS-Dwarfs}
\label{sec:cos-dwarfs}

The COS-Dwarfs survey comprises HST/COS spectra
of quasars whose sightlines penetrate the CGM of
$z \sim 0$ dwarf galaxies \citep{bordoloi14}.
We provide the full dataset, binned at 3~pixels.

\subsection{UVpSM4}
\label{sec:hstmetals}

\cite{ctp+10,cpt+11} compiled and processed all of the medium
resolution ($R \sim 2000$) and high resolution
($R > 20000$) UV spectra available in 2010 
(i.e.\ prior to the SM4 servicing mission)
to study metal-line absorption in the $z<1$ IGM.
From {\it HST}, these were primarily STIS and
GHRS datasets.  The authors also included
supporing data from the Far-Ultraviolet Spectrographic 
Explorer (FUSE).
All of these fluxed spectra are ingested and we
also include the \cite{ctp+10} continuum models.

\subsection{HSTQSO}
\label{sec:hstqso}

\cite{ribaudo11} and \cite{neeleman+16}
compiled nearly the entire set of UV spectra of 
quasars and AGN available in the {\it HST} archive
to survey for Lyman limit and damped \lya\ systems.
This includes the Faint Object Spectrometer dataset
compiled by \cite{bechtold02} and data from GHRS,
STIS and COS (Lehner et al., in prep.).
The data group includes 360 spectra from COS,
339 spectra with FOS, and 205 spectra taken
with STIS.  For the COS spectra, we have ingested
the files provided by the Hubble Space Legacy 
Archive\footnote{https://archive.stsci.edu/hst/spectral\_legacy/}
(HSLA).
Note that there is overlap in spectra
between this data group and
the other {\it HST} datasets listed above, although
each was processed separately.

\subsection{HST\_z2}
\label{sec:hstz2}

\cite{omeara11,omeara13} obtained slitless grism
(with the Wide Field Camera 3) and prism (with the
Advanced Camera for Science) spectra with {\it HST}
of optically bright quasars at $z \sim 2.5$
to survey LLS at $z \sim 2$.  
We have ingested their entire dataset.

\section{Architecture of the {\it igmspec} Data File}
\label{sec:arch}

\begin{figure}
\includegraphics[width=6in]{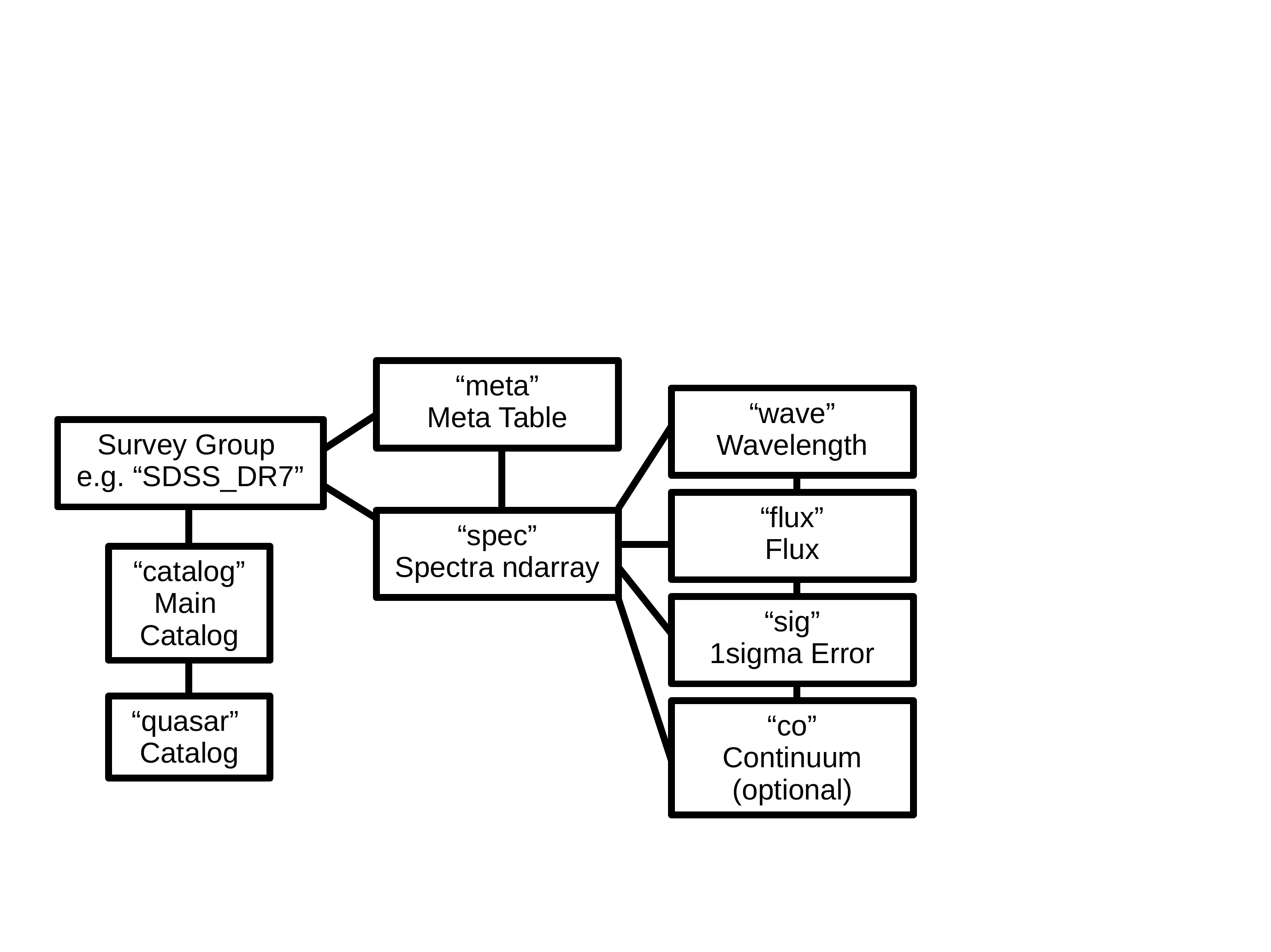}
\caption{Schematic describing the architecture of 
the {\it igmspec} database.
}
\label{fig:arch}
\end{figure}

The {\it igmspec} database is provided as a single HDF5 file
(IGMspec\_DB\_v02.hdf5)
containing the source catalog, a separate quasar catalog,
and the spectra with their meta data.  
The HDF5 format enables rapid access to the data without
reading the entire database into memory.  
Figure~\ref{fig:arch} illustrates the 
architecture of the HDF5 file.
It was built using Python's {\it h5py} package
and software within the {\it igmspec, specdb, linetools}
and {\it astropy} packages.

Each dataset comprises an HDF5 Group
with a {\it meta} Dataset and a {\it spec} Dataset.
The former is first generated as an 
{\tt astropy} Table, with one row per
spectrum, and then converted into an HDF5 object.
The latter is an {\tt numpy.ndarray} 
with dtype names `wave', `flux', and `sig' for the
wavelength, flux, and $1\sigma$ error arrays.
The wavelength values are stored as float64 data type
and the rest are float32.
Many of the datasets also include 
an estimate of the source continuum which is
recorded in the `co' column.

The {\it igmspec} database file may be
downloaded using the spec\_get\_igmspec script in {\it specdb}
without installing the {\it igmspec} repository.

\section{Software}
\label{sec:software}

The {\it igmspec} repository\footnote{https://github.com/specdb/igmspec} 
includes a set of Python modules and scripts to build the
{\it igmspec} database.  
These are not intended to be used by the general community.
Access to the {\it igmspec}
database file is provided within the 
{\it specdb} repository.  This section summarizes the key
software in the {\it specdb} repository with emphasis
on accessing and using {\it igmspec}. 

\subsection{Downloading {\it igmspec}}

The {\it specdb} repository includes a simple script for retrieving
a copy of any of its public databases.  This includes {\it igmspec};
the corresponding Python script is specdb\_get\_igmspec.
This script uses a {\tt wget} call to the URL of the database file.
A related script, specdb\_chk, summarizes the contents and creation data
of any {\it specdb} database file.

\subsection{Interfacing with the Database in Python}

The {\it specdb} repository\footnote{https://github.com/specdb/specdb}
includes software developed in Python
for interacting with a {\it specdb} database file.  The primary
object is the SpecDB class and one may use a child IgmSpec for the
{\it igmspec} database.
After instantiating this object, one may query the source catalog,
the spectra meta data, and retrieve spectra into memory.
Details are provided in the online {\it specdb} 
documentation\footnote{http://specdb.readthedocs.io/en/latest/}
and here is a summary:

\begin{itemize}
\item {\it Querying the source catalog:}  The SpecDB class instantiates
a QueryCatalog class to load and then query the source catalog.  The
majority of methods use position on the sky or a user-input set of
coordinates in the queries.  Documentation is given here\footnote{http://specdb.readthedocs.io/en/latest/catalog.html\#querying-the-source-catalog}
and the repository includes an iPython Notebook\footnote{https://github.com/specdb/specdb/blob/master/docs/nb/Query\_Catalog.ipynb}
with examples.

\item {\it Querying the spectral meta data:}  Each spectrum in the
database has associated meta data which may be queried.  The interface
is the InterfaceGroup class which reads the data into an astropy Table
and performs queries.  The relevant documentation is found 
here\footnote{http://specdb.readthedocs.io/en/latest/meta.html}
and the repository includes an iPython Notebook\footnote{https://github.com/specdb/specdb/blob/master/docs/nb/Query\_Meta.ipynb}
with examples.

\item {\it Retrieving spectra:} The InterfaceGroup class contains the
low-level routines for spectral retrieval, but we recommend the higher-level
methods within the SpecDB class.  These methods return the data packaged
within an XSpectrum1D\footnote{http://linetools.readthedocs.io/en/latest/xspectrum1d.html} 
object from the {\it linetools} package.
Documentation is described 
here\footnote{http://specdb.readthedocs.io/en/latest/spectra.html}
and examples are located in an iPython 
Notebook\footnote{https://github.com/specdb/specdb/blob/master/docs/nb/Retrieving\_Spectra.ipynb}
in the repository.

\end{itemize}

We also refer the reader to an iPython 
Notebook\footnote{https://github.com/specdb/specdb/blob/master/docs/nb/Examples\_with\_igmspec.ipynb}
for examples with {\it igmspec}.
Lastly, there is a set of command-line 
scripts\footnote{http://specdb.readthedocs.io/en/latest/scripts.html}
which enable the user to access the spectra without
launching an explicit Python session.

\subsection{SSA Interfacing}

Although we primarily intend for the {\it igmspec} database
to be downloaded and then
accessed and manipulated locally 
(the HDF5 file is only $\approx 25$\,Gb),
the {\it specdb} software includes an SSAInterface object for 
performing standard SSA queries on a {\it specdb} database file.
The current implementation query 
takes POS (position) and SIZE (angular radius) 
inputs and returns an SSA v1.1 compliant VOTable. 
One can also perform a FORMAT=METADATA query.
Referring to v2.0 of the IVOA spectral data model (dated
2016-09-28),  all of the mandatory fields of the model
are provided except: (1) Char.SpatialAxis.Coverage.Bounds.Extent,
as the aperture is not precisely known for all spectra in
{\it igmspec};
(2) Char.TimeAxis.Coverage.Bounds.Extent, because
the total exposure time is not always recorded for 
spectra coadded across multiple exposures.
The current interface also does not include software
to distribute individual spectra.  If community pressure is
sufficient, we will stage {\it igmspec} within an online 
SSA service.

\subsection{Building a {\it specdb} Database}

The {\it specdb} repository includes Python methods 
to generate new {\it specdb} databases.
The basic requirements are:
 (i) files of the spectra that can be read by the spectra
 tools in {\it linetools};
 (ii) listings of the spectral meta data that provide at least
 the default information in Table~\ref{tab:meta_spec};
 (iii) source astrometry and redshifts.
For the latter, one can interface with the {\it igmspec} database.
The software also includes a means to add IVOA descriptors
for meeting spectral data model 2.0.  

\section{Concluding Remarks}
\label{sec:end}

With the {\it igmspec} database, it is our goal to provide (nearly)
all of the published spectral datasets that effectively
probe the IGM.  Hopefully, this effort will
enable new, unforeseen research on the IGM as well
as the greater diffusion of otherwise difficult-to-access spectral
datasets.  

In v03 of
{\it igmspec} (expected release date is $\sim 6$ months from
this publication), we plan to include at least the following:
(1) additional near-IR spectroscopy;
(2) radio absorption spectra \citep[e.g. 21\,cm;][]{kanekar+14};
(3) galaxy spectra probing the IGM \citep[e.g.][]{rpk+10}, 
and
(4) spectroscopy of GRB afterglows \citep[e.g.][]{fjp+09}.
Community members interested in guiding the future development
of {\it igmspec} are encouraged to contribute via github
(https://github.com/specdb/igmspec).

To enable IGM cross-correlation analyses with galaxies
and the large-scale structures they trace,
we intend the future release of {\it exgalspec}.
This database will have -- at
the minimum -- a catalog of (nearly) all spectroscopically
confirmed galaxies and, where feasible, their associated
spectra.  See https://github.com/specdb/exgalspec
to contribute to that effort.

\section{Acknowledgments}

J. X. P. is partially supported by NSF grant AST-1412981.
We thank many individuals who helped with the construction
of the database:
John O'Meara, Gabor Worseck, Joe Ribaudo, Marcel Neeleman,
Jason Tumlinson, Rongmon Bordoloi, Jessica Werk,
Kathy Cooksey, S. Croom, K.-G. Lee, Guangtun Zhu,
Adam Myers, Joseph Hennawi,  and Marc Rafelski.

JXP thanks the Pacific Research Platform (PRP) whose support
helps host the files for {\it igmspec}.  The PRP is funded by
NSF Project \#ACI-1541349 to
PI Larry Smarr at the University of California, San Diego.




\end{document}